# Does Plasmodium falciparum have an Achilles' heel?

## Liao Y. Chen[1†]

[1]Department of Physics, University of Texas at San Antonio, One UTSA Circle

San Antonio, Texas 78249 USA

Email: Liao.Chen@utsa.edu

**ABSTRACT**

**Background**: Plasmodium falciparum is the parasite that causes the most severe form of malaria responsible for nearly a million deaths a year. Currently, science has been established about its cellular structures, its metabolic processes, and even the molecular structures of its intrinsic membrane proteins responsible for transporting water, nutrient, and waste molecules across the parasite plasma membrane (PPM).

**Presentation of the hypothesis:** I hypothesize that Plasmodium falciparum has an Achilles' heel that can be attacked with erythritol, the well-known sweetener that is classified as generally safe. This hypothesis is based on the molecular structure of the parasite's membrane and the quantitative mechanics of how erythritol interacts with the multi-functional channel protein expressed in the PPM. Most organisms have in their cell membrane two types of water-channel proteins: aquaporins to maintain hydro-homeostasis across the membrane and aquaglyceroporins to uptake glycerols etc. In contrast, P. falciparum has only one type of such proteins---the multi-functional aquaglyceroporin (PfAQP) expressed in the PPM---to do both jobs. Moreover, the parasite also uses PfAQP to excrete its metabolic wastes (ammonia included) produced at a very high rate in the blood stage. This extremely high efficiency of the bug using one protein for multiple essential tasks makes the parasite fatally vulnerable. Erythritol in the blood stream can kill the parasite by clogging up its PfAQP channel that needs to be open for maintaining hydro-homeostasis and for excreting toxic wastes across the bug's PPM.

**Testing the hypothesis:** *In vitro* tests are to measure the growth/death rate of P. falciparum in blood with various erythritol concentrations. *In vivo* experiments are to administer groups of infected mice with various doses of erythritol and monitor the parasite growth levels from blood samples drawn from each group. *Clinic trials* can be performed to observe the added effects of administering to patients erythritol along with the known drugs because erythritol was classified as a safe food ingredient.

**Implications of the hypothesis:** If proven true, erythritol will cure the most severe form of malaria without significant side effects.

**Background**

Does Plasmodium falciparum---the parasite that causes the most severe form of malaria leading to the death of a child every 60 second [1]---have an Achilles' heel that can be exploited therapeutically? The answer seems to be yes even though the bug enjoys three layers of protection by the red cell membrane (RCM), the parasitorphorous vacuole membrane (PVM), and the parasite plasma membrane (PPM). On the basis of the molecular structure of the bug's PPM and the quantitative mechanics of its major intrinsic membrane protein interacting with erythritol, I hypothesize that erythritol [2]--- the generally considered safe sugar substitute [3]---could be the weapon needed to deliver a fatal attack on the heel of the parasite that is responsible for about a million deaths a year.

Currently, all malaria drugs used or researched for are intricate compounds that have strong side effects and induce drug-resistance in the parasites.[4-17] Researchers have not considered the possibility that the benign erythritol can kill or impede the growth of P. falciparum. In fact, erythritol can move across the cell membrane through the conducting pore of aquaglyceroporins just like glycerol which is a nutrient for the parasite's growth [18] and up taken into the P. falciparum cell across, consecutively, the RCM, the PVM, and the PPM.[19, 20] The possibility of erythritol being an inhibitor of P. falciparum is visible only in the light of the parasite's molecular structure available recently in the literature[20-22] and the quantitative determination[23-25] of how erythritol interacts with the channel protein in the PPM illustrated in Figure 1.

**Presentation of the hypothesis**

The hypothesis is that erythritol in the blood stream will kill P. falciparum or, at the least, impede the parasite's growth in the human blood stage. The rationale is the following:

(1) Recent *in vitro* studies concluded that P. falciparum has only one type of water-channel proteins, aquaporins (AQPs), expressed in its plasma membrane---Plasmodium Aquaporin (PfAQP).[21, 22] This is in contrast with other organisms that all have two types of water-channel proteins in their cell membranes---aquaporins dedicated for water transport and aquaglycero-porins that conduct water, glycerol, ammonia and other small solutes.[26] P. falciparum relies on PfAQP for glycerol uptake, water conduction, and excretion of its metabolic wastes produced at high speed in the blood-stage.[27] The waste molecules, if not excreted, would intoxicate the bug itself within minutes.[20]

(2) Needless to say, a living cell needs its waste-excretion channel unclogged in order to not be intoxicated by its own metabolism. Less obvious but equally true, a living cell needs its water channel unclogged as well to maintain hydro-homeostasis across its cell membrane. Otherwise, it will not be able to survive the severe osmotic stress during renal circulation.

(3) The conducting channel of PfAQP, like all other aquaporins, is single file in nature.[22] As shown in Fig.1, water and solute molecules line up in single file throughout the conducting pore. When an erythritol molecule dwells inside the pore instead of passing through the channel instantly, it clogs it up, inhibiting the transport of water, glycerol, and waste molecules across the cell membrane.

(4) The chemical-potential profile of erythritol in PfAQP tells us that there will always be a erythritol dwelling inside the channel when erythritol is administered to a patient at a rather low dosage. The binding affinity of erythritol to PfAQP corresponds to a half-maximal inhibitory concentration of $IC_{50}$=786 nM [25]. Therefore, when the erythritol concentration in the blood is in the high μM range, the binding site inside PfAQP will be saturated with erythritol. The PfAQP channel will be clogged up nearly all the time for transporting water, glycerol, and waste molecules. Consequently, P. falciparum will lose its ability to excrete wastes and to survive osmotic stress.

Now, what the advantages does a permeant of the channel protein have over a total channel blocker which were suggested to be ineffective [28]? First, a total blocker of the PfAQP channel probably inhibits other members of the aquaglyceroporin subfamily including AQP3 and AQP9 that are expressed in the red cell and other human cells. Indiscriminate inhibition of aquaglyceroporins in all red cells alone (infested with P. falciparum or not) will certainly cause undesirable side effects, not counting other cells. In contrast, erythritol permeates aquaglyceroporins in the RCM, PVM, and PPM. Consequently, equilibrium of erythritol concentration is expected between the parasite's cytosol and the serum outside the red cells, assuming that the parasite does not consume erythritol. If the parasite does metabolize erythritol, depending on the rate of consumption, a concentration gradient would exist in the cytoplasmic direction. Even in such a case, the PfAQP pore will still be occupied and thus occluded by an erythritol with a probability of nearly 100% if the extracellular concentration of erythritol is far above the $IC_{50}$ of 786 nM. Moreover, erythritol taken orally enters quickly into the blood stream and has a long biological half-life [29]. Therefore, any amount of erythritol would practically inhibit the highly efficient functions of PfAQP in facilitating transport of water, ammonia, urea, and glycerol. Finally, erythritol is known to be safe, not causing side effects.

**Testing the hypothesis**

*In vitro* experiments: Adding erythritol to non-immune human whole blood, prepare 10 culture media with erythritol concentrations ranging from 0·0 mM to 0·9 mM by increment of 0·1 mM, nine media from 1·0 mM to 9·0 mM by increment of 1·0 mM, and 10 media from 10 mM to 100 mM by increment of 10 mM. Measure the growth and death rates of P. falciparum in these 29 media. Repeat these experiments multiple times to minimize the statistical uncertainties.

*In vivo* experiments: Divide mice infected with P. falciparum into 20 groups. Give each group a different daily dose of erythritol. Group One, 0 mg/kg, Group Two, 10 mg/kg, Group Three, 20 mg/kg, ……, Group 10, 90 mg/kg, Group 11, 100 mg/kg, Group 12, 200 mg/kg, ……, and Group 20, 1000 mg/kg. Observe the survival rate and time of each group. During the course of time, take blood samples and measure the parasite growth in each group. Repeat the experiments to minimize the statistical uncertainties. Note that even the highest dosage suggested here does not have any adverse effects to the physiology of humans.

**Implications of the hypothesis**

Finally, a physician can safely administer a range of doses of erythritol along with other anti-malaria drugs to controlled patient groups and observe the added effectiveness due to erythritol at various dosages. If the hypothesis is proven true, we will soon see the eradication of the most severe form of malaria.

**List of abbreviations**

AQP---Aquaporin/Aquaglyceroporin

PfAQP---Plasmodium falciparum aquaglyceroporin

PPM---Parasite plasma membrane

PVM---Parasitorphorous vacuole membrane

RCM---Red cell membrane

**Acknowledgements**

The author acknowledges support from the NIH (Grant #GM084834) and the Texas Advanced Computing Center.

**FIGURES**

Figure 1. PfAQP channel occluded (left) and open (right) for water/solute permeation. The luminal residues of the protein are shown in wireframes colored by residue types, the waters in red (oxygen) and white (hydrogen) balls, and the erythritol in black balls (hydrogen, oxygen, and carbon). Not all luminal residues of PfAQP are shown for the purpose of fully exposing the waters and erythritol that line up in single file inside the conducting pore. The z-axis points from the extracellular to the cytoplasmic bulk. Graphics rendered with Virtual Molecular Dynamics.[30]

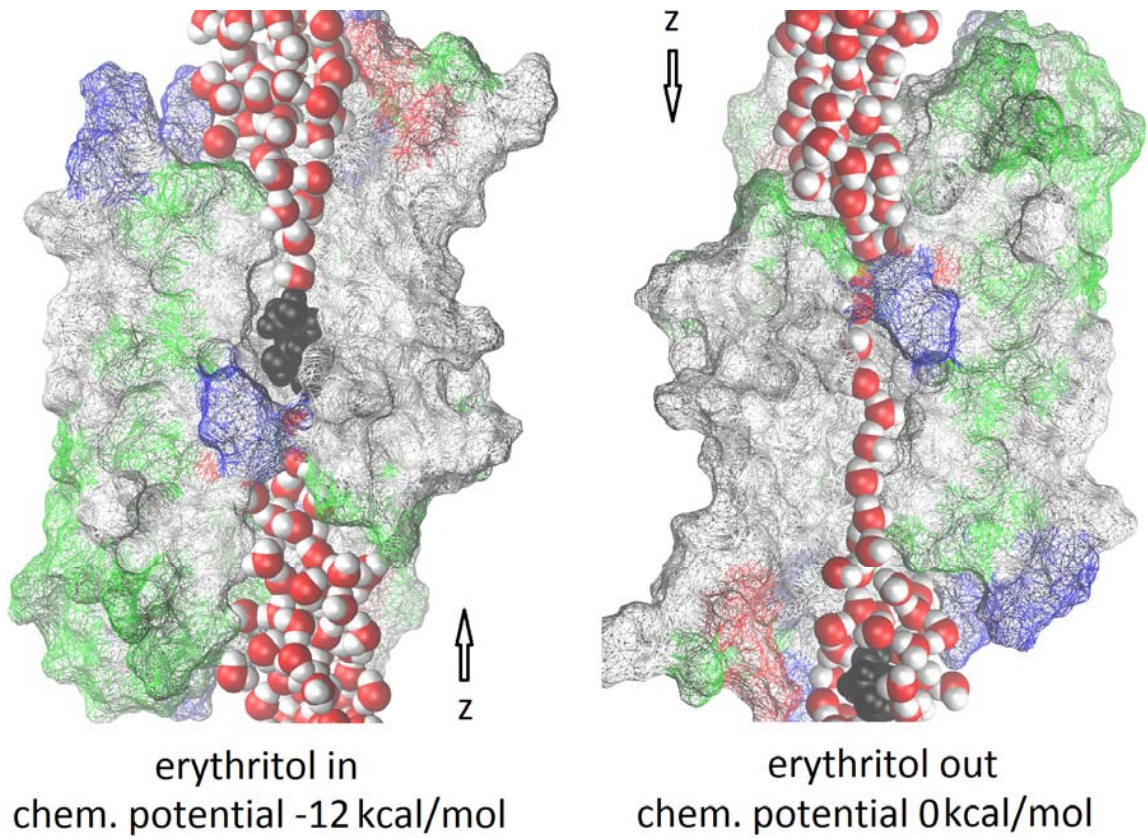

erythritol in
chem. potential -12 kcal/mol

erythritol out
chem. potential 0 kcal/mol